\documentclass[sigchi]{acmart}

\renewcommand\footnotetextcopyrightpermission[1]{} 

\usepackage{booktabs} 

\usepackage{float}

\usepackage{amsmath}
\usepackage{caption}

\DeclareMathOperator*{\argmin3}{arg\,min3}

\setcopyright{none}

\acmConference{Submitted for review}
\acmYear{2018}
\begin{document}
\title{Live Emoji: Semantic Emotional Expressiveness of 2D Live Animation}

\author{Zhenjie Zhao}
\affiliation{%
  \institution{Hong Kong University of Science and Technology}
  \city{Hong Kong}
}
\email{zzhaoao@cse.ust.hk}

\begin{abstract}
    Live animation of 2D characters has recently become a popular way for storytelling,
    and has potential application scenarios like tele-present agents or robots.
    As an extension of human-human communication, there is a need for augmenting the
    emotional communication experience of live animation.
    In this paper, we explore the emotional expressiveness issue of 2D live animation.
    In particular,
    we propose a descriptive emotion command model to
    bind a triggering action, the semantic meaning, psychology measurements, and behaviors of an emotional expression.
    Based on the model, we designed and implemented a proof-of-concept 2D live animation system, where
    a novel visual programming tool for editing the behaviors of 2D digital characters,
    and an emotion command recommendation algorithm are proposed.
    Through a user evaluation, we showcase the usability of our system and
    its potential for boosting creativity and enhancing the emotional communication experience.
\end{abstract}

%
%




\maketitle

\section{Introduction}


Live animation, as a new form of storytelling, opens the possibility of online communication
through controlling a digital character.
For example, through the usage of Adobe Character Animator \cite{AdobeCharacterAnimator},
users can manipulate the behaviors of a puppet in real-time during online chatting.
As previous research works show that emotional expressiveness can bring more engaged experience
to audiences \cite{Akazue:2016:ETS:2858036.2858307,Sonderegger:2016:AAA:2858036.2858365},
equipping 2D live animation with emotional tools can potentially fulfill
online communication tasks, e.g., news broadcasting, interacting with students or children
remotely, etc., better.

However, expressing emotional states of 2D live animation is challenging 
\cite{Sonderegger:2016:AAA:2858036.2858365,Willett:2017:TAS:3126594.3126596,ma2012guidelines}. 
Generally, facial expressions of digital puppets are the most salient channel for expressing emotions
\cite{Sonderegger:2016:AAA:2858036.2858365}.
Existing approaches, such as expression transfer \cite{Cao:2016:RFA:2897824.2925873}
and swapping artworks \cite{Willett:2017:TAS:3126594.3126596},
are still difficult or not easy for users to communicate full range emotional states reliably.
For expression transfer, namely, capturing users' facial expressions and
deforming the digital puppet accordingly, it is difficult for users to express different deliberate
emotions. 
For swapping artworks that
changes different puppet faces according to different
emotional states,
it is also challenging to design artworks with different emotions that can be 
recognized reliably, especially for subtle emotions \cite{ma2012guidelines}.
In addition, users might need to customize emotions to
express specific semantic meanings \cite{repurposeemoji}.
The basic categories of emotions \cite{scherer2005emotions},
i.e., happy, sad, fearful, angry, surprised, disgusted, are not enough to express
secondary emotions or emotional experiences in different contexts precisely 
\cite{zagalo2005emotional}. For instance,
the emotional state of a person riding a roller coaster is hard to be classified into
any one category.
To express secondary emotions, dimensional emotion models are needed,
such as the circumplex model \cite{posner2005circumplex},
the Geneva emotion model \cite{scherer2005emotions}, and so on, which describe
the geometric emotion space under two or more channels \cite{zagalo2005emotional}.
However, designing different puppet faces with different secondary emotions and swapping them during
live animation is still a challenging task, and even not possible if the story is not planned
beforehand, i.e., no prior contextual information.


The diverse emoji and sticks in current online communication tools provide a novel perspective
to deal with the emotional expressiveness problem in live animation.
With the usage of
emoji and stickers, senders can express the emotion feeling of a text message better,
and receivers can also feel the sentiment of the message more saliently.
Popular chatting softwares, e.g., Wechat \footnote{\url{weixin.qq.com}},
QQ \footnote{\url{im.qq.com}}, and so on,
build in powerful emoji functions to provide a rich emotional chatting experience,
which is very welcomed by users \cite{Zhou:2017:GTH:3025453.3025800}.
Specifically, stickers exaggerate the traditional emoji, 
and are extremely popular among the younger generation in China \cite{Zhou:2017:GTH:3025453.3025800}.
Compared to traditional emojis,
stickers are usually created by modifying real photos,
cartoons, with exaggerated facial expressions, background animated texture patterns,
and illustrated texts to enhance emotion feeling.

Using emotional expressing ways of stickers
can potentially enhance the emotional expressiveness ability of
live animation, which leads to better
storytelling and communication experience. In this paper, we explore how to help users make
rich emoji with expected semantic meanings, e.g., happy, sad, etc.,
and control customized emoji in real-time through live animation.
In particular, we designed and implemented an easy-to-use 2D live animation system to
help users express emotional states more efficiently.
To represent diverse secondary emotions,
following the results of
Sonderegger \emph{et. al} \cite{Sonderegger:2016:AAA:2858036.2858365},
we used a dimensional emotional model, and
decomposed the emotional state into valence and arousal levels.
We collected a dataset of popular stickers, and identified
three main components for expressing emotional meanings, i.e., facial expressions,
background texture patterns, and texts. Finally,
we built a visual programming tool that allows users to combine different
elements of the three components to create a rich emoji easily and intuitively.

To conceptualize the emotional expressiveness during live animation,
we propose an \emph{emotion command} model that binds a triggering action, 
the semantic meaning of an emotion,
psychology measurement metrics \cite{scherer2005emotions}, and the behaviors of a digital puppet together.
The model makes it convenient for users to connect
different expected semantic meanings with a particular puppet's behaviors
during live animation, and also helps simplify the design of our live animation system.
In addition, with the bound psychology measurements,
we can also build recommendation algorithms to recommend particular emotion commands
during live animation. Through a preliminary user evaluation,
we show that our system can help users build interesting emoji, and
express emotional states easily during live animation.
The contributions of this paper can be summarized as:

\begin{enumerate}
    \item Building on the dimensional emotion model,
    we propose an \emph{emotion command} model to conceptualize emotional expressiveness
    of live animation. In addition, leveraging the procedural animation concept, we designed and implemented
    a visual programming tool to edit an emotion command.

    \item Targeting online communication scenarios, 
    we designed and implemented a proof-of-concept 2D live animation system.

    \item We conducted a preliminary user evaluation to explore the usability of our system.
\end{enumerate}

\section{Related Works}

\subsection{Live Animation}

Controlling a puppet to tell stories has been a long time theatrical practice.
Previous works of digitalizing live animation
include simulating Chinese shadow puppetry
\cite{Lu:2011:SCC:1978942.1979221},
controlling marionettes in a Virtural Reality (VR) environment
\cite{Sakashita:2017:YPE:3126594.3126608}, and so on.
The recent development of face tracking techniques brings new design forms of
digital live animation tools, e.g., the software Adobe Character Animator.
In \cite{Willett:2017:TAS:3126594.3126596}, researchers propose a recommendation
triggering strategy using touch screens, which significantly decreases the perceptional
workload during live animation.
In \cite{Willett:2017:SMP:3126594.3126641},
by the usage of physical simulation, researchers add secondary motion to
a primary motion to make more lively visual effects.

Previous works usually focus on how to align the motion trajectories of
performers and the puppets, but how to augment the movements of
performers to express rich emotional states correctly is more challenging.
The introduction of secondary motion \cite{Willett:2017:SMP:3126594.3126641}
provides promising results in this direction.
However it is still problematic for users to express different
secondary emotions mentioned previously.

\subsection{Emotional Representation}

Emotional experience is important for better social communication
\cite{dolan2002emotion}, and plays an important role
in human-human interaction, human-robot interaction, and human-computer interaction.
Allowing an agent to express diverse emotional states could engages its users better,
and has attracted many research efforts recently.

For emotional representation,
one crucial question is how to define emotion itself.
In general, there are three different points of view, i.e.,
discrete categories, e.g., Ekman's six basic
emotions \cite{ekman1971universals}, the dimensional model, e.g.,
the circumplex model \cite{posner2005circumplex}, and the appraisal-based model,
e.g., the Geneva emotion model \cite{scherer2005emotions}, which can be seen
as an extension of dimensional models. Discrete categories
are shown to be cross-cultural, and easy for humans to recognize,
but suffer from describing secondary emotions. Therefore,
to represent diverse emotional states, dimensional models are expected.
Another problem is which communication channel should be used to convey emotions.
Works on displaying emotions include
different sensory channels, for instance,
temperature \cite{Tewell:2017:HTD:3025453.3025844,Wilson:2016:HUC:2858036.2858205},
color \cite{Bartram:2017:ACV:3025453.3026041},
embodied emotional feedback \cite{Hassib:2017:EAE:3025453.3025953},
vibration \cite{Bucci:2017:SCC:3025453.3025774},
visualization and animation \cite{Sonderegger:2016:AAA:2858036.2858365},
and the fusion of multiple signals
\cite{Wilson:2017:MCT:3025453.3025614}, and so on.

Inspired by the popularity of stickers
\cite{Zhou:2017:GTH:3025453.3025800}, our work explores representing emotions by dimensional models
through facial expressions, 
animation of background texture patterns, 
and explicit illustration with texts.
In addition, to enable users to express diverse semantic meanings of emotional states during live animation,
an easy-to-use tool for composing different emoji is needed.

%

\subsection{Visual Programming}

As suggested in \cite{Sarracino:2017:USI:3025453.3025467},
using programming tools can help create more complex animation effects.
However, programming is usually a challenging task for non-professionals
\cite{Sarracino:2017:USI:3025453.3025467}.
Researchers have explored different approaches to facilitate this procedure, e.g., direct manipulation on graphics
\cite{Sarracino:2017:USI:3025453.3025467,Kazi:2014:DBL:2556288.2556987}, visual programming tools, etc.
For example, in \cite{Kazi:2014:DBL:2556288.2556987},
\emph{Draco} is proposed, which allows users to
program the expected motion trajectories by drawing. Similarly,
a physical illustration tool in \cite{Sarracino:2017:USI:3025453.3025467}
allows users to program the relationships
of different objects by connecting them with a mouse.

Using visual programming tools is another alternative to make programming accessible
by novices, e.g., the visual programming language \emph{Scratch}
\cite{Resnick:2009:SP:1592761.1592779}.
Among many visual programming tools, the behavior tree model
is a simplified version for editing behaviors of digital characters, which is
widely used in the game industry.
However, the general design of a behavior tree model is still not flexible enough for
the context of emotional expressiveness of live animation. And for novices,
picking up concepts in behavior tree models is also
challenging. More care should be taken to tailor a suitable behavior tree model
for the usage of emotional expressiveness of live animation.

The remaining of the paper is organized as follows. We first present the proposed emotion 
command model. And then we introduce the interface design of our live animation system,
following the implementation details, including a visualization tool,
a visual programming tool, the default live animation function, and
a recommendation algorithm. In the following, we present the initial use case study
and result analysis. Finally, we discuss the limitations of this work and potential futher 
directions.

\section{Emotion Command Model}

To conceptualize emotional expressiveness of 2D live animation, we propose
an \emph{emotion command} model.
An emotion command combines a triggering action, the semantic meaning of
an emotion, different measurement metrics, e.g., valence,
arousal, etc., and the animated behaviors together.
By triggering an emotion command, a particular emotional behavior can be presented.
The conceptualization of emotional expressiveness helps simplify
the system design procedure. In addition, we can also
help users build their customized emotional communication vocabulary.
Finally, with the model, it is also manipulatable to design algorithms for recommending the expected
emotion commands during live animation.

An emotion command can be represented as a five element tuple:
\begin{equation*}
  (d, s, v, a, b),
\end{equation*}
where $d \in \mathcal{D}$ is a triggering action of a particular device (keyboard in this paper),
$s \in \mathcal{S}$ is the semantic meaning of the emotion command given by users,
$v \in \mathbb{R}$ and $a \in \mathbb{R}$ are
the valence (happiness) level and arousal (activation) level reported by users respectively,
$b \in \mathcal{B}$ is the behavior actions of the puppet triggered
by $d$.

To construct $b$ with diverse expressiveness using basic components,
we collected 216 stickers from QQ and Wechat.
In particular, we used the emic approach \cite{headland1990emics}, and treated ourselves
as participants in several chatting groups and peer-to-peer chatting processes.
We collected stickers that can express a strong emotional feeling during chatting.
Some stickers are animated while some are not.
Several examples of the dataset are shown in Table \ref{sticker_dataset}.
To code the dataset,
the authors had regular meetings and discussions, 
and finally identified three main components for conveying rich emotional feeling:
facial expression $\mathcal{F}$, background texture pattern $\mathcal{T}$,
and text $\mathcal{O}$.
Therefore, a behavior action is a tuple: $b=(f^*, t^*, o^*)$,
where $*$ denotes $b$ can contain zero, one, or more sub-action $f$, $t$, $o$,
$f \in \mathcal{F}$, $t \in \mathcal{T}$, and
$o \in \mathcal{O}$.

\begin{table}[h]
  \centering
  \begin{tabular}{ccc}
    \includegraphics[width=0.28\columnwidth]{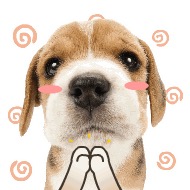}
    & \includegraphics[width=0.28\columnwidth]{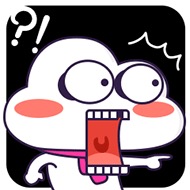}
    & \includegraphics[width=0.28\columnwidth]{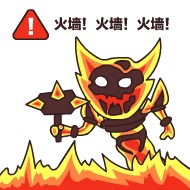} \\
    \includegraphics[width=0.28\columnwidth]{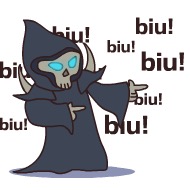}
    & \includegraphics[width=0.28\columnwidth]{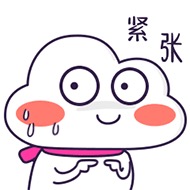}
    & \includegraphics[width=0.28\columnwidth]{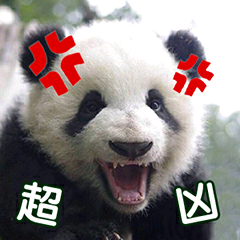} \\
  \end{tabular}
  \caption{Several examples of our collected emotional sticker dataset.}
  \label{sticker_dataset}
\end{table}

\vspace{-20pt}


\section{User Interface}

We designed and implemented a proof-of-concept 2D live animation system \emph{Live Emoji} to
demonstrate the \emph{emotion command} model.
The general workflow of our system is that users first edit
a corpus of emotion commands through
a visual programming tool, connect to a remote partner,
control a telepresence puppet through live animation, and then trigger emotion commands
to enhance the emotional communication experience.

We denote the two partners communicating with each other as performer and audience,
where the performer is the side that users control the puppet and trigger the emotion commands,
and the audience is the side that users watch and interact with a telepresence puppet.

\begin{figure}
  \centering
  \includegraphics[width=1.0\columnwidth]{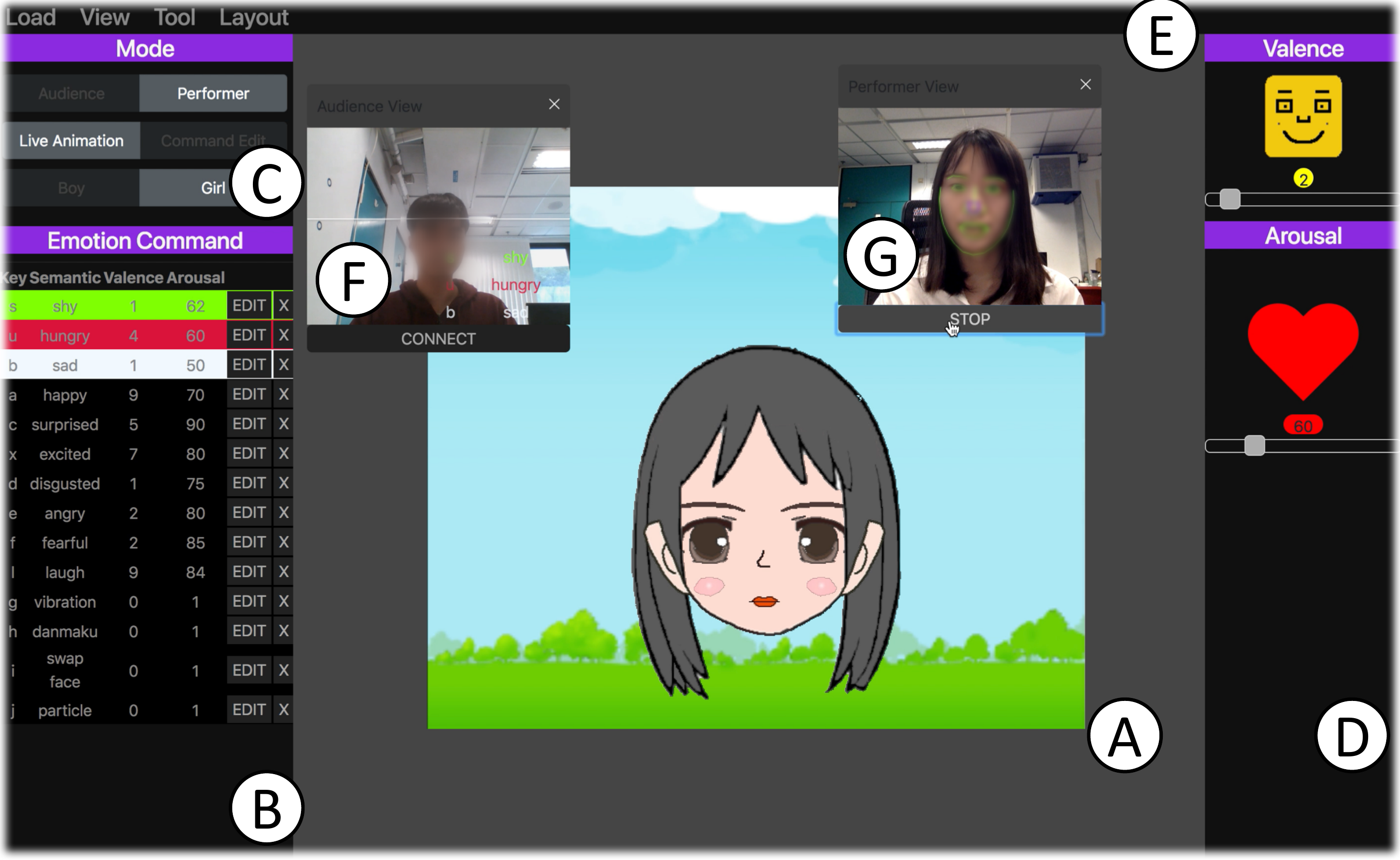}
  \caption{The performer side interface of the proposed live animation system.}
  \label{Interface of our system}
\end{figure}

The interface of the performer side is shown in Figure \ref{Interface of our system},
which contains the main canvas (A) for displaying the animated puppet,
the emotion command view (B) for collecting all the emotion commands
edited by users, the mode view (C) for choosing different usage mode, i.e.,
the layout of the interface (audience, performer), usage stage (live animation,
editing emotion commands),
and character selection (boy, girl),
the emotion view (D) for visualizing the valence and arousal
levels for helping users edit the emotion command,
the menu bar (E) for other helpful functions, e.g.,
toggling the performer view (F) that displays users themselves,
the audience view (G) that displays the remote partner,
the visual programming view (Figure \ref{The visual programming interface})
to edit emotion commands, changing the background image, and so on.

The audience side only contains the main canvas (A), the optional audience view (F) for displaying
the remote partner, and the menu bar (E).

\subsection{Workflow}

\subsubsection{Editing Emotion Command}

To edit an emotion command $(d, s, v, a, b)$, users need to
input the bound key $d$ and the semantic meaning $s$, e.g.,
happy, sad, etc. The emotion view (D) provides a \emph{visualization tool} to assist users to
self-evaluate the valence level $v$ and arousal level $a$
of the expected semantic meaning.
\emph{The visual programming tool} allows users to visually edit
multiple behaviors $b$ shown in the main canvas (A),
including facial expressions, background texture patterns, and texts.

\subsubsection{Live Animation}

During live animation, we built in
a \emph{default live animation} function, where the system detects users' facial information for
controlling basic states of the digital puppet, e.g., open or close eyes, face position, etc.
Users can trigger an emotion command by pressing the key $d$ to display the animated behaviors $b$.
In addition, with a corpus of emotion commands pre-edited by users,
the system can estimate users' valence and arousal levels during live animation,
and recommend several emotion commands.

\section{System Implementation}
We implemented the system using Javascript.
The scene graph is organized with Threejs \footnote{\url{https://threejs.org}}. Apart from face tracking,
for other face related algorithms,
i.e., facial expression recognition, eye blink detection,
face landmark detection, mouth open or closed detection,
we adopted existing libraries
through a trial and error procedure.
The communication between two partners
was implemented with the technology WebRTC \url{webrtc.org},
and our system can be used with an ordinary network setting.

\subsection{Character Representation}

To simplify the system, we only provide pre-set characters
with head portraits.
Two professional artists were invited to help us
design the characters,
including a boy character and a girl character, as shown in Table
\ref{The designed characters used in our live animation system.}.
The face portrait of each character is discretized as:

\begin{equation*}
    \begin{split}
        \Gamma=\{\textbf{static}, \textbf{eye}^L_i, \textbf{eye}^R_i, \textbf{eyebrow}^L_i, \textbf{eyebrow}^R_i, \\
        \textbf{mouth}_i, \textbf{nose}_i\},
    \end{split}
\end{equation*}

\noindent where $\textbf{static}$ denotes static decorative parts that are not changed during
live animation, including left and right ears, front face, and hair,
$L$ denotes left, $R$ denotes right, $i$ is the emotional state, and

$i \in \{\textbf{happy},\textbf{sad},\textbf{fearful},\textbf{angry},
\textbf{surprised},\textbf{disgusted}\}$.

\begin{table}[h]
  \centering
  \begin{tabular}{c}
    \includegraphics[width=22em]{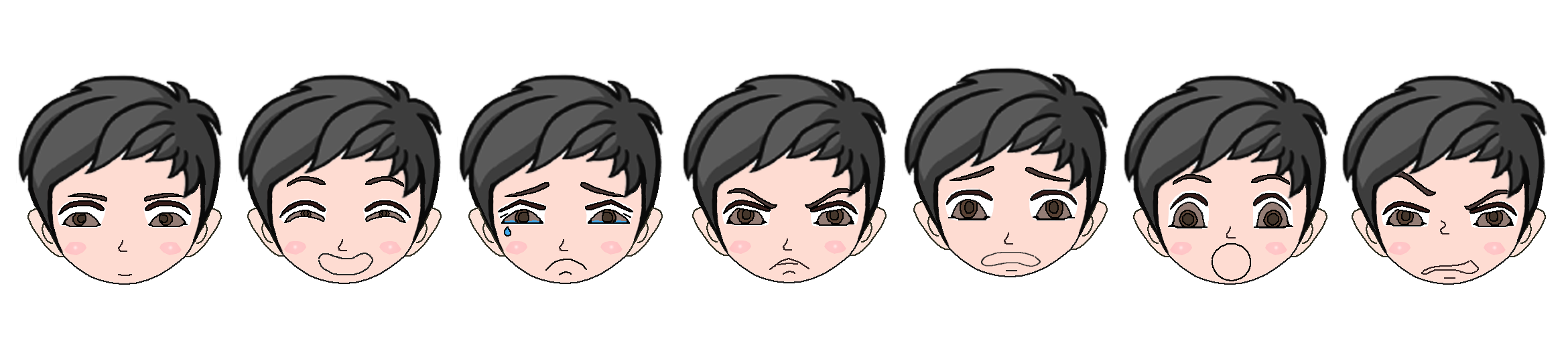} \\
    \includegraphics[width=22em]{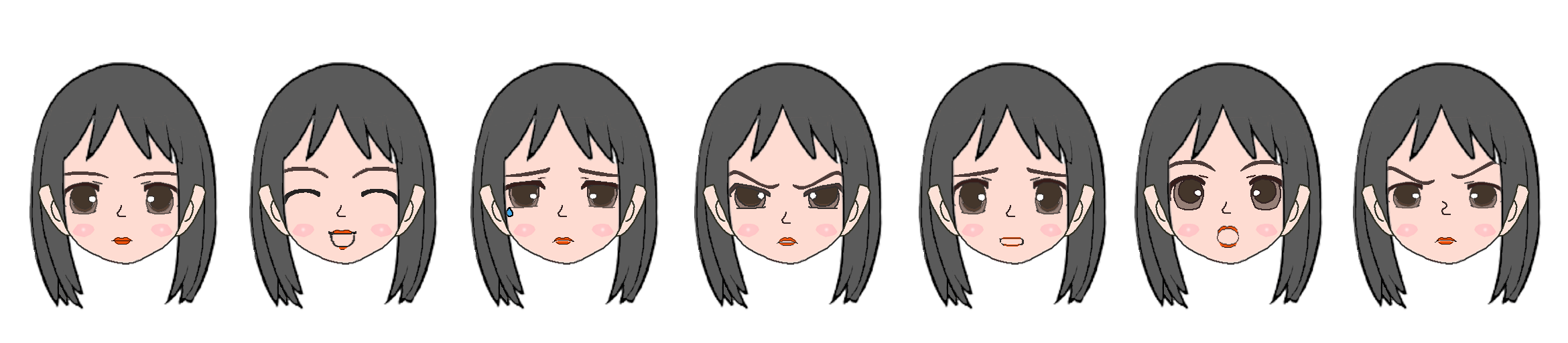}
  \end{tabular}
  \caption{The designed characters (1st row: boy, 2nd row: girl) used in \emph{Live Emoji}.
  From left to right, the expressions are: neutral, happy, sad, angry, fearful, surprised, disgusted.}
  \label{The designed characters used in our live animation system.}
\end{table}

The controllable parameters of $\Gamma$ are $(x, y, i, \zeta_l, \zeta_r, \xi)$,
where $x$ and $y$ are the horizontal and vertical positions, $i$ is the emotion state,
$\zeta_l$ indicates the close state of left eye, i.e., if $\zeta_l=1$, left eye is closed,
similarly, $\zeta_r$ is the close state of right eye, $\xi$ is the open state of mouth,
and if $\xi=1$, mouth is open. During live animation, the system estimated the state of users
to control $x$, $y$, $\zeta_l$, $\zeta_r$ automatically, and $i$ is triggered by
emotion commands to change different emotion artworks of the character.

To enable the reliable recognition of different facial expressions,
we adopted the guidelines in \cite{ma2012guidelines} for designing facial features of six basic emotions.
Although the six emotions are not enough for users to describe various secondary emotional states,
for facial expressions except the six basic ones, it is usually hard to be recognized
reliably \cite{ma2012guidelines}. Therefore, it might be confusing
for users to choose different facial expressions.
In addition, according to the circumplex model \cite{posner2005circumplex},
the six basic emotions can cover most parts of
the geometric distribution uniformly, which means
any emotions including the secondary emotions mentioned previously
can be approximated by one of the six emotions.
Referring to the using practice of emoji,
our design consideration is to rely on the combination of the three different channels,
i.e., facial expression, background animation, and text,
to express subtle emotions.
The feedback from users also indicates the approach is meaningful.
In addition, we found that the way of speaking emotion meanings straightly with animated texts
is especially useful.


\subsection{Visualization Tool for Emotion Evaluation }

To edit an emotion command, users need to evaluate the valence and arousal value of the semantic
emotion meaning, which are used for building the recommendation algorithm.
We designed a visualization tool to assist users to self report the expected values, as shown in
Table \ref{The emotion module.}. For both panels, users can drag the slider to
change the values, and the corresponding visualization and numerical results are updated.
A professional artist was invited to draw the facial expressions of different valence levels referring to
the 9 scale Self-Assessment-Manikin (SAM) \cite{bradley1994measuring} (1-9, 5 by default).
Inspired by \cite{Sonderegger:2016:AAA:2858036.2858365}, where animated background can convey
different levels of arousal meaning better, we used an animated heart to indicate different arousal levels.
The numerical range is set to be $50 \sim 90$ (70 by default) to
reflect the common sense of normal heart rate, where higher heart rate indicates
higher arousal level, and lower heart rate indicates lower arousal level.

\begin{table}[h]
  \centering
  \begin{tabular}{cccc}
    \includegraphics[width=6em]{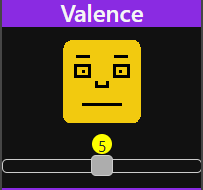} &
    \includegraphics[width=6em]{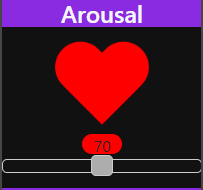}
  \end{tabular}
  \caption{The visualization tools for self reporting valence (left) and arousal (right) level.}
  \label{The emotion module.}
\end{table}

\vspace{-20pt}

\subsection{Visual Programming Tool for Behavior Editing}

\begin{figure*}
  \centering
  \includegraphics[width=1.8\columnwidth]{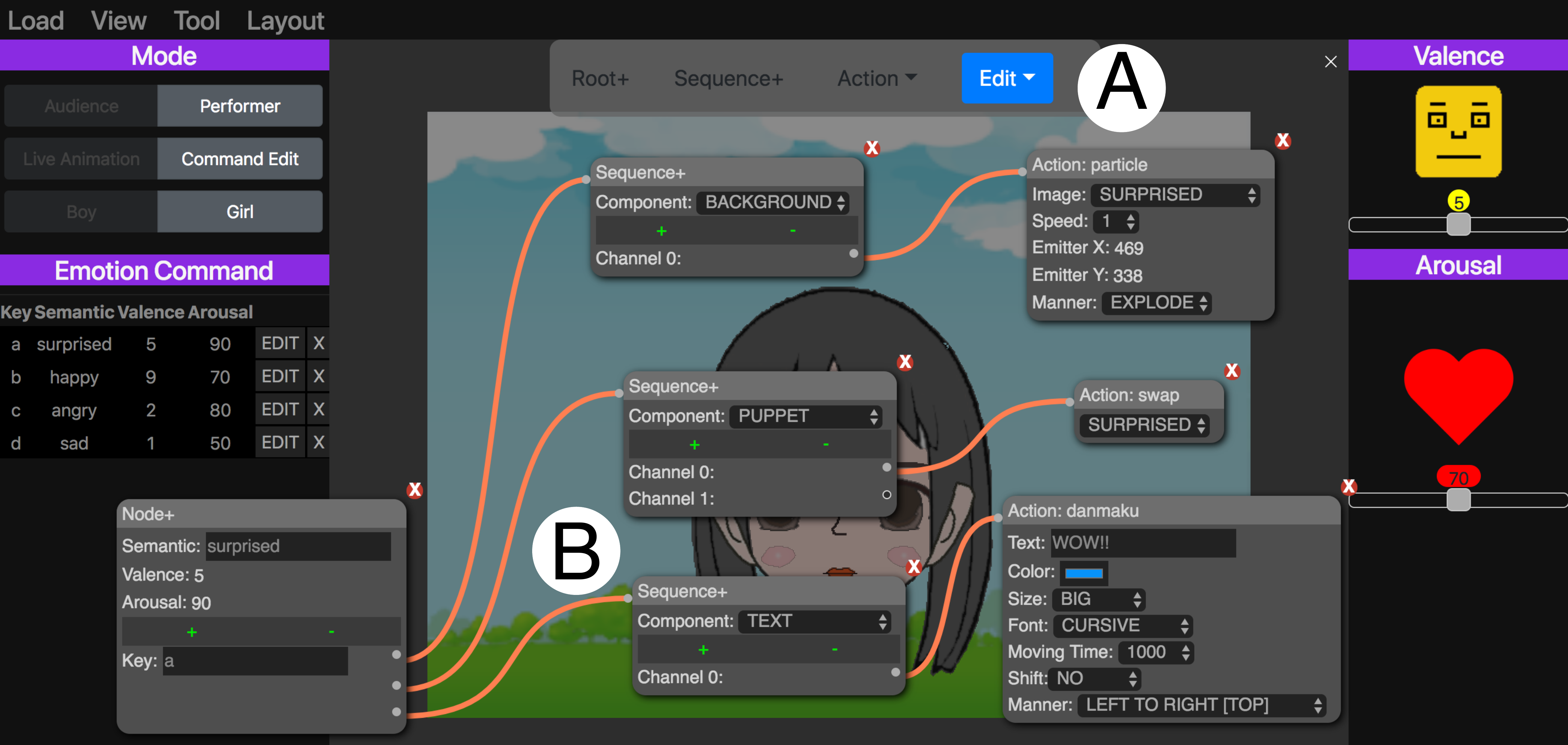}
  \caption{The visual programming interface, including the navigation bar (A) and the main canvas (B).}
  \label{The visual programming interface}
\end{figure*}

We modified a standard behavior tree model (BT) for building the visual programming tool,
which provides users an easy-to-use tool for composing an emotion command.

A behavior tree model is a directed acyclic graph $\mathcal{G}(\mathcal{V}, \mathcal{E})$ with
$|\mathcal{V}|$ nodes and $|\mathcal{E}|$ edges.
The child-less nodes are called \emph{leaf},
and the parent-less nodes are called \emph{root}.
Each node in a BT,
is one of seven possible types: root, four non-leaf control-flow node types, i.e.,
selector, sequence, parallel, and decorator, and two leaf execution node types, i.e.,
action and condition. For a digital puppet, a \emph{shot} of behaviors can be
composed by one root node, with several control-flow nodes, and several execution nodes,
with the structure:

\begin{equation*}
\text{root} \rightarrow \text{control} \rightarrow \text{execution},
\end{equation*}

The root node is to indicate the target puppet, the control node is to
control the logic of the behaviors, and the execution node is the real actions of the digital puppet.
In addition, users can concatenate several \emph{shot}s of behavior together to compose more complicated
behavior actions.

Although behavior tree is widely used in game industry, enabling novices to pick up
the aforementioned concepts is still difficult. We modified several definition rules
to better fit our emotional state manipulation task and make it intuitive for novices.

\subsubsection{Design Rationale}

With the \emph{emotion command} model and our emoji dataset,
we designed the modified behavior tree with the following rationales:

\begin{enumerate}
  \item Users should have the flexibility to add any combination of the three basic components, i.e.,
  facial expression, background animation, and text, which are commonly seen in online chatting emoji.
  \item Users can bind any number of the behavior actions to a particular component.
  \item Some actions can be used only for specific components, e.g., the face swapping action for
  the face component. However, to give users more degree of freedom to connect component nodes and
  action nodes and simplify the behavior tree model,
  it is desirable to abstract several common behavior actions of the three different components, e.g.,
  vibration.
\end{enumerate}

\subsubsection{Modified Behavior Tree}

To meet the design requirements, we included only three types of nodes in our modified behavior tree model,
root+, sequence+, and action.
The visual programming tool of an emotion command has the following structure:

\begin{equation*}
\text{root+} \rightarrow \text{sequence+} \rightarrow \text{action}.
\end{equation*}

The interface of the visual programming tool is shown in Figure \ref{The visual programming interface},
which includes a navigation bar (A) for adding different nodes (root+, sequence+, action)
and editing the command (e.g., new, save, import, export), and the main canvas (B) for visual programming the
emotion command.
Root+ node is a parametric node for binding the semantic meaning $s$, valence $v$, arousal $a$, and key $d$.
One sequence+ node corresponds to a target component, i.e., facial expression, background animation,
text, and it is used to connect different action nodes for that component.
The behaviors $b$ are composed by sequence+ and action nodes together.
The connection between the root+ node to the sequence+ node allows
users to compose an emoji by selecting any combination of the
components, and the connection between the sequence+ and action node
provides the flexibility to decorate various behaviors of a component.

\subsubsection{Action Node}

We designed five different action nodes for the proof-of-concept system,
but with the general emotion command model, it is easy to add additional
behavior actions.

\textbf{Danmaku Action}.
The danmaku action is inspired by the text comment animation behavior on online video websites,
and it is only used for text component.
Users can choose to edit the text, font size, color,
and the animation manner.
Similar to danmaku, the animated texts can be moved from right to left or left to right of the screen.
We also built in a shift option to add displacement of the text position,
which allows users to place several animated texts on the screen.
Although the action is simple, we found it is very helpful for
expressing emotional meanings straightly through our user evaluation.

\textbf{Swap Action}.
The swap action node is used for swapping the facial expressions of the digital puppet. Users
can choose one from the six basic emotion templates
shown in Figure \ref{The designed characters used in our live animation system.}.

\begin{figure}
    \centering
    \includegraphics[width=1.0\columnwidth]{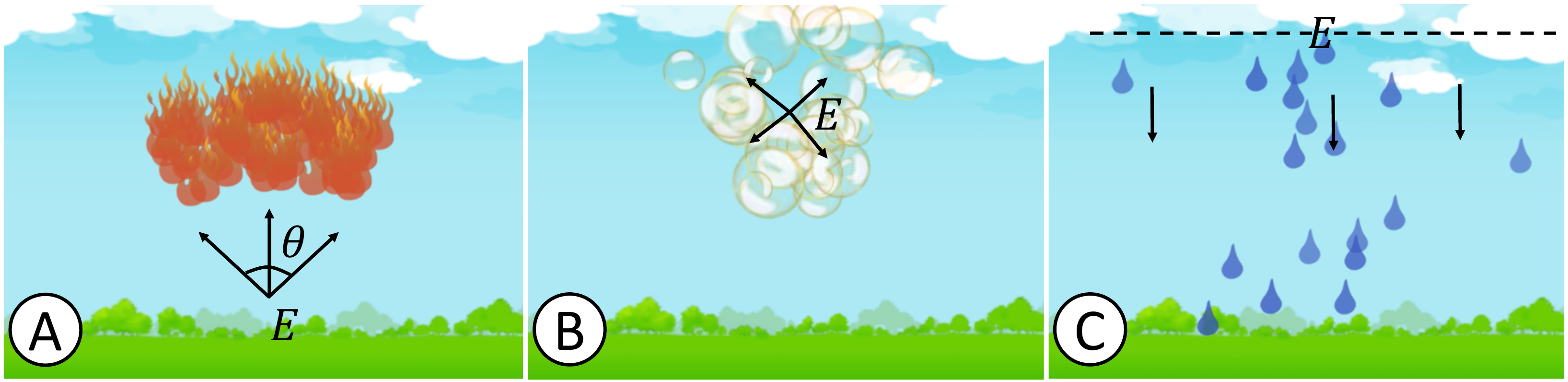}
    \caption{The motion patterns of particle node, including A) jet, B) exploding, C) rain.}
    \label{The motion patterns of particle node}
\end{figure}

\textbf{Particle Action}.
The particle action is used for creating the particle system of the background component.
Users can choose the image texture, the emitter point, the motion pattern (jet, exploding, rain),
and the speed of each particle.
As shown in Table
\ref{The collection of image textures designed by artists for decorating the particle system for background animation.},
We invited artists to design several image textures with six basic emotional meanings,
and colorized them referring to the affective color design in \cite{Bartram:2017:ACV:3025453.3026041}.
Three typical motion patterns from existing cartoon background effects were implemented,
as shown in Figure \ref{The motion patterns of particle node}.
For jet motion, we uniformly emit particles within an angle $\theta$ around the emitting point
with initial velocity,
and particles are not influenced by other forces during moving.
For exploding motion, particles are produced randomly around the emitting point $E$ with initial velocity,
and influenced by gravity. For rain motion, we extend two segments with equal length around the
emitting point horizontally, and emit particle randomly along the segments with zero
initial velocity, and the particles are influenced by gravity.
For all three motion patterns, we gradually make the particle appearance transparent,
and add small noises on the initial velocity to make more diverse effects.
Again, it is easy to add more texture images and motion
patterns for augmenting the current functionality.

\begin{table}
  \centering
  \begin{tabular}{cccccc}
    \includegraphics[width=2.6em]{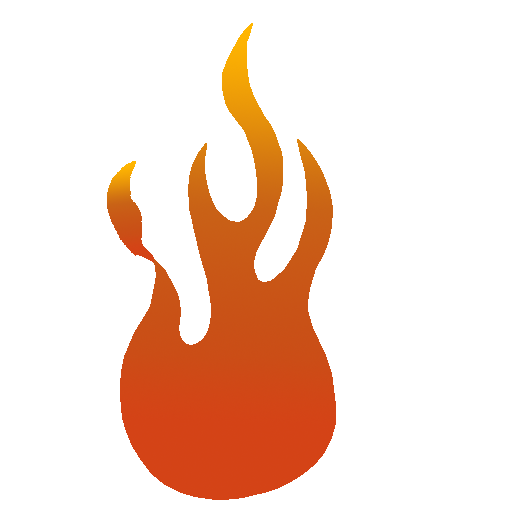} &
    \includegraphics[width=2.6em]{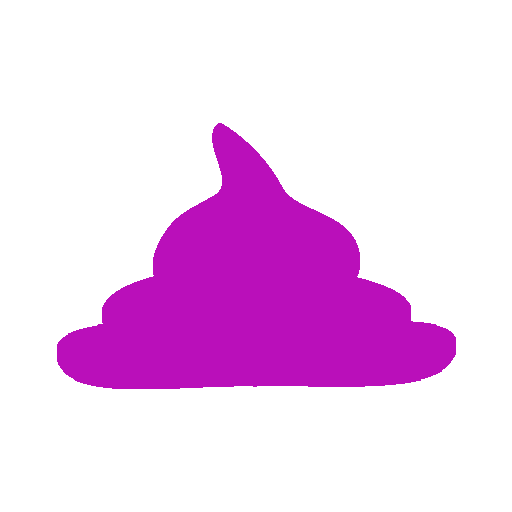} &
    \includegraphics[width=2.6em]{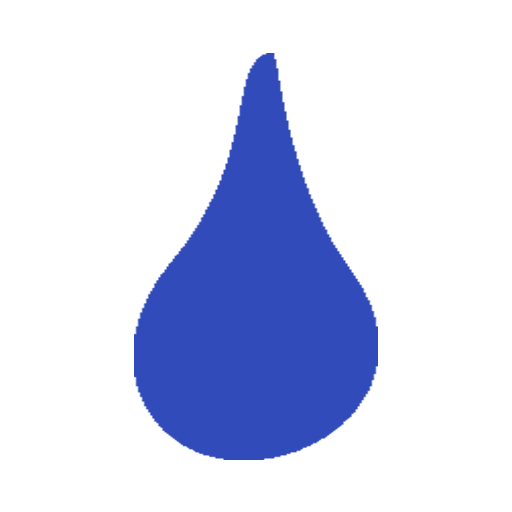} &

    \includegraphics[width=2.6em]{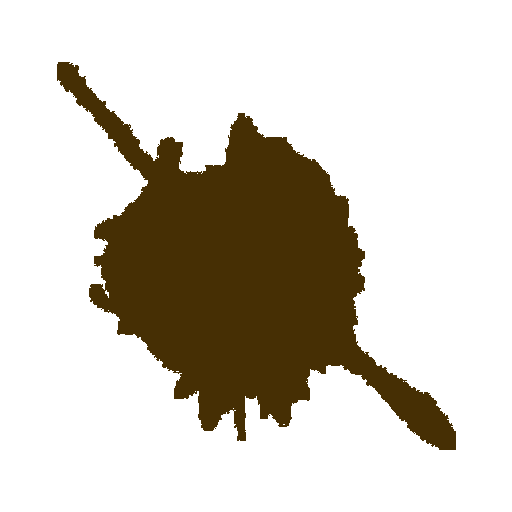} &
    \includegraphics[width=2.6em]{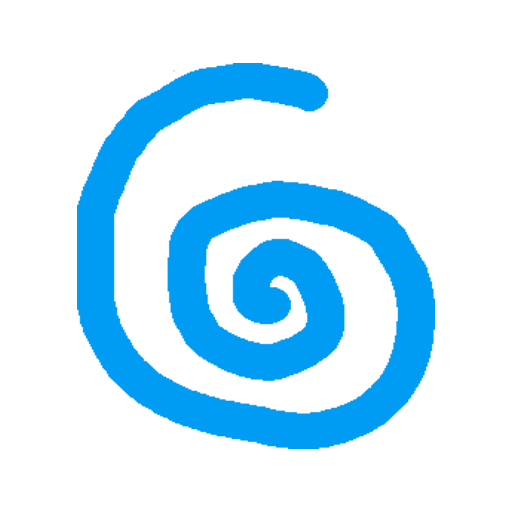} &
    \includegraphics[width=2.6em]{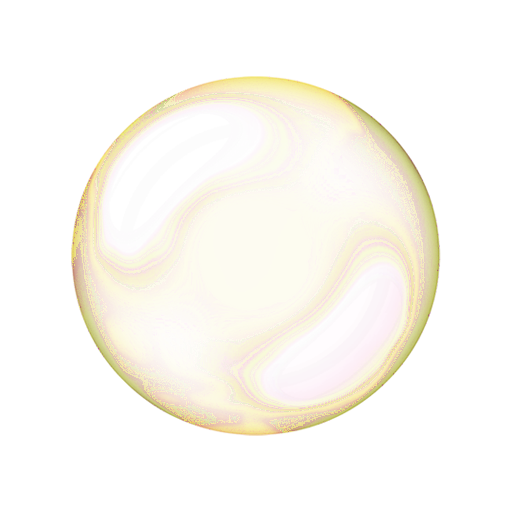} \\
    angry & disgusted & sad & fearful & surprised & happy \\
  \end{tabular}
  \caption{The collection of image textures designed by artists for decorating the particle system
  for background animation.}
  \label{The collection of image textures designed by artists for decorating the particle system for background animation.}
\end{table}

\textbf{Vibration Action}.
To demonstrate the concept that we can abstract common actions of the three components,
we designed a vibration action node that uses one degree-of-freedom vibration behavior for expressing
emotions. The vibration action can be connected to all the three components. When a
vibration node is triggered, its corresponding component will shake around its position.
We allow users to select different levels of frequency and amplitude.
The interpolation between endpoints is done by \emph{tween.js} \footnote{\url{https://github.com/tweenjs/tween.js}}.

\textbf{Sound Action}. We also designed a sound action node that can be connected
to the facial expression component.
Several human emotional sounds, e.g., laugh, scream, etc., can be chosen in
this node. The sound effect files were downloaded from Free Sound Effects
\footnote{\url{https://www.freesoundeffects.com/free-sounds/human-sound-effects-10037/180/tot_sold/20/10}}.

\textbf{Background Image}. Apart from the five action nodes, users can also change the background
image that presents different stages for the puppet. Several images from \emph{Graphic Mama}
\footnote{\url{https://graphicmama.com/blog/free-cartoon-backgrounds}} were downloaded
for usage during the evaluation session.

\subsection{Default Live Animation}

To provide a live communication
experience with the puppet, we bound the controllable parameters $x, y, \zeta_l, \zeta_r, \xi$
to the detected states of users by default.
When toggling the live animation function, these states of the puppet will follow the
detected ones of users.
For eye, mouth open / closed detection, and the face position tracking,
we used a face landmark detection library
\emph{clmtrackr} \footnote{\url{https://github.com/auduno/clmtrackr}} to locate users' eyes,
mouth, and face.
\emph{clmtrackr} can outputs 71 landmark coordinate positions of the face,
and the result was robust enough for our application during testing.

\subsubsection{Open / closed eye detection}
From face landmark detection, we can get the eye regions.
We used \emph{keras}
\footnote{\url{https://github.com/keras-team/keras}} to train a convolution neural network on the
CEW blink detection dataset \footnote{\url{parnec.nuaa.edu.cn/xtan/data/ClosedEyeDatabases.html}}
for classifying the open and closed eyes,
and imported the trained mode using \emph{keras.js}
\footnote{\url{https://github.com/transcranial/keras-js}} to our system.
The validation accuracy of the model is 96.17 \%, which is enough for our system.

\subsubsection{Open / closed mouth detection}
We applied a simple way by detecting the volume of
users' voice within a buffering size,
If the volume is larger than 0.02, then the mouth is open.
Otherwise, it is closed.

\subsubsection{Face Tracking}
To alleviate face position drift and jitter effects, face tracking is necessary.
In addition, to limit the
computation intensive deep learning based algorithms, i.e., open / closed eye detection,
emotion recognition (in the following section),
a relatively low refresh rate should be given, and thus the tracking
procedure can also alleviate the drift problem of low detection rate.
In particular, we designed a particle filter for tracking the face.
The dynamics of face position is modeled as:
\begin{equation*}
  \left\{\begin{matrix}
  \frac{dx}{dt} = \varepsilon_1
  \\
  \frac{dy}{dt} = \varepsilon_2
  \end{matrix}\right.,
\end{equation*}
where $t$ denotes time, and $\varepsilon_1 \sim \mathcal{N}(0, \delta^2)$,
$\varepsilon_2 \sim \mathcal{N}(0, \delta^2)$ are two different Gaussian random
variables. The hidden variable is $\boldsymbol{h}=[x, y, dx/dt, dy/dt]^T$,
and the observable variable is $\boldsymbol{o}=[x, y]^T$.
The measurement of the face position $m$ is the average value of all
landmark positions $l_i$, $m=\frac{1}{N} \sum_{i=1}^N l_i$.
We adopted a standard
condensation particle filter, and used 500 particles to track
the position of the face. After testing different illumination conditions, we
were satisfied with the tracking result.

\subsection{Emotion Command Recommendation}

Once the corpus of emotion commands becomes big, searching the expected
emotion command during live animation could be difficult. Therefore,
recommending potential suggestions according to users' spontaneous emotions
is necessary.
To build a recommendation algorithm, we need to first estimate the
valence $V$ and arousal level $A$ during live animation.
However, detecting users' spontaneous emotional states
is still a difficult task \cite{4468714}. For building the live animation system,
we explored an experience-oriented approach, where we manually
tuned the detected signals to match the mental values self reported by users. And we got two
empirical formulas for estimating the valence and arousal levels separately.
Based on our initial user evaluation, users were generally satisfied with the results.

\subsubsection{Valence Estimation}
We adopted a categorized emotion recognition engine
\footnote{\url{https://github.com/oarriaga/face_classification}}
to estimate the valence level.
And the trained model was imported into our system using \emph{keras.js}.
The emotion recognition engine can return the probabilities of six basic
emotions or the neutral emotion $P_i$, where

$i \in \{\textbf{happy},\textbf{sad},\textbf{fearful},\textbf{angry},
\textbf{surprised},\textbf{disgusted}\}$.

To encode the probability to the valence value $V$ ($V \in [1,9]$),
we separate emotions of the circumplex model as positive (right semi-circle)
and negative (left semi-circle).
For the largest detected probability $P_{max}=max\{ p_i \}$,
if the corresponding emotion is positive, the valence is calculated
as $V=\left \lfloor P_{max} \times 4 \right \rfloor + 5$, otherwise,
$V=\left \lfloor P_{max} \times 4 \right \rfloor + 1$.

\subsubsection{Arousal Estimation}
\begin{figure}
    \centering
    \includegraphics[width=0.7\columnwidth]{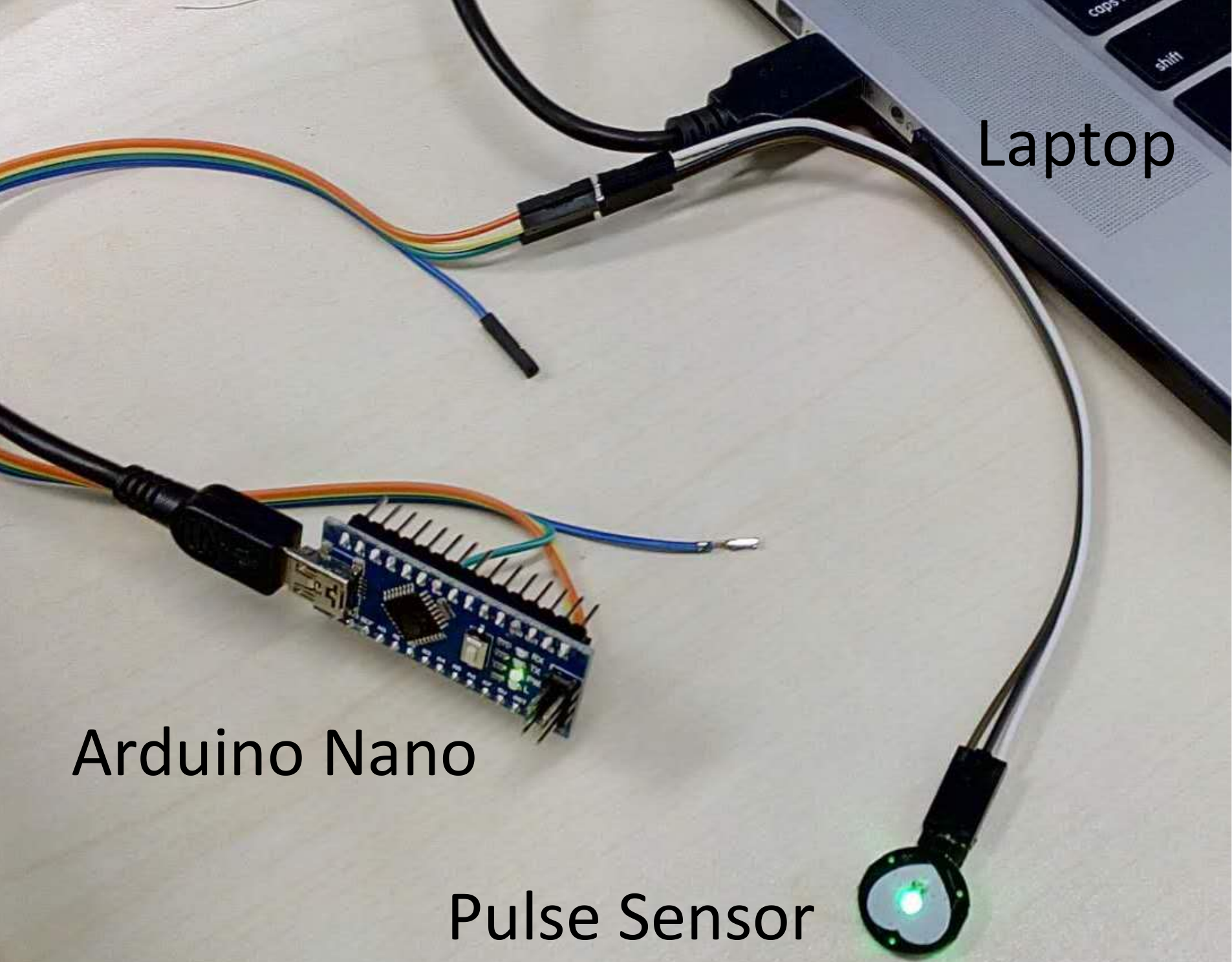}
    \caption{The device for measuring heart rate.}
    \label{The device for measuring heart rate.}
\end{figure}

We used the heart rate to approximate the arousal level, and adopted
a low cost solution to measure it. As shown in Figure
\ref{The device for measuring heart rate.}, a pulse sensor
\footnote{\url{https://pulsesensor.com}}
is connected the laptop through an Arduino nano
\footnote{\url{https://store.arduino.cc/usa/arduino-nano}},
and we used the Johnny-Five robotics framework
\footnote{\url{johnny-five.io}} for
processing the measured signal. Beats per minute (BPM) is
approximated by counting the frequency of peak values in real-time.
During experiments, we found that the measured BMP was not aligned
with the mental range of arousal values ($50-90$) well. Therefore,
we correct the arousal level with BPM as $A=70 + \gamma(BPM-60)$, where $\gamma$
is a scale factor parameter ($\gamma=1.05$).

\subsubsection{Recommendation Algorithm}
By coupling the two dimensional coordinates,
we can get the input emotion feature as $E=(V, A)$.
For the pre-edited $N$ emotion commands, each consists an emotion measurement
$e_i = (v_i, a_i)$, where $i=1,\ldots,N$.
Based on the circumplex emotion model \cite{scherer2005emotions},
a natural way for measuring the similarity of two emotional states
is to calculate their Euclidean distance.
However, during experiments, due to the noises of valence and arousal
estimations, the un-alignment of self reported valence and arousal with the
estimated results during live animation, we found that
that users were generally not satisfied with the recommended proposals.
We went through several versions of designing the recommendation algorithms,
and finally decided to recommend the emotion command according to positive
and negative emotions separately:
\begin{equation*}
  R3=\left\{\begin{matrix}
  \argmin3\limits_{em_i \in EM_{positive}} d(em_i, V, A), & V \ge 5\\
  \argmin3\limits_{em_i \in EM_{negative}} d(em_i, V, A), & V < 5
\end{matrix}\right.,
\end{equation*}
where $R3$ is the collection of three recommended emotion commands,
$\argmin3$ is the operator of finding arguments of the top three minimal
distances $d(em_i, V, A)$, $em_i$ is the $i$-th emotion command,
$EM_{positive}$ is the collection of all positive emotion commands, i.e.,
emotion command with $v \ge 5$, and
$EM_{negative}$ is the collection of all negative emotion commands, i.e.,
emotion command with $v < 5$.
The distance is calculated as the weighted Euclidean measurement:
\begin{equation*}
   d_i = \sqrt{ ||V-v_i||^2 + \alpha ||A-a_i||^2},
\end{equation*}
where $||\cdot||^2$ is the normalized distance ($0 \sim 1$),
$\alpha$ is the weight coefficient between the valence and arousal
levels.
During our testing, we found that the estimated arousal did not change too much to
reflect the real arousal value of the expected emotion. Therefore,
we use a smaller weight ($\alpha=0.5$) to discount the influence of arousal
for computing the distance.

As shown in Table \ref{The recommendation interface.},
during live animation, we encode the top three recommended commands with
different colors, and place them at the top of the emotion command view (left).
In addition, to make it convenient for navigating during communication,
the recommended commands are also displayed in the audience view (right).

\begin{table}[h]
  \centering
  \begin{tabular}{cccc}
    \includegraphics[width=12em]{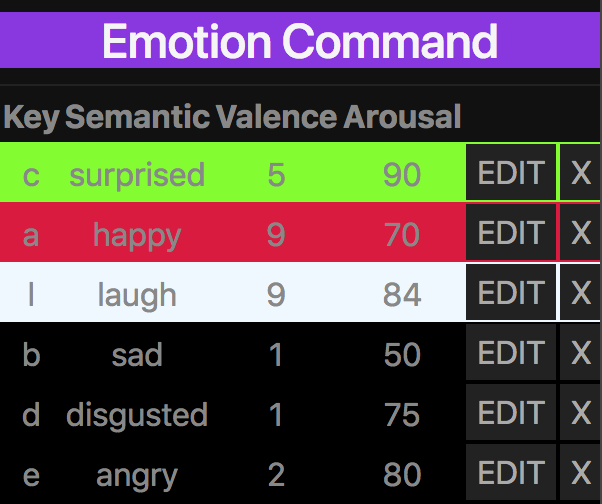} &
    \includegraphics[width=10em]{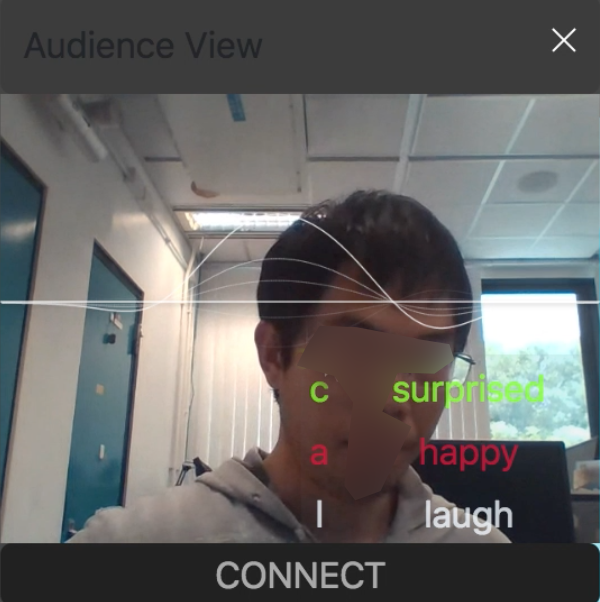}
  \end{tabular}
  \caption{The recommendation interface (left),
  and the top three commands are also shown in the audience view (right).}
  \label{The recommendation interface.}
\end{table}

\vspace{-30pt}

\section{Initial User Evaluation}

In this paper, we designed and implemented \emph{Live Emoji}
for concept proof of the proposed emotion command model,
which conceptualizes the emotional expressiveness problem of 2D live animation.
To further explore the usefulness of the system and collected feedback for improvement,
we conducted an initial user evaluation.
Instead of formally evaluating the usability of our system
or comparing with other approaches,
our study only served as an exploration of the emotion command
concept and the potential application scenario from a user's point of view.
Therefore, we only consider the evaluation from the performer side.
Before conducting the user evaluation,
we first went through the functions of our system in a group meeting,
and adjusted the corresponding features according to the comments to
make the system more user friendly.

\subsection{Participants}

Six participants (P1-P6) took part in our user evaluation.
Two participants (P2, P4) are professional graphic illustrators with
moderate to good drawing skills, and the remaining have little
experience on graphics design and drawing.
All participants are frequent users of emoji.

\subsection{Experiment Setting}
We prepared six basic emotion commands
with semantic meanings happy, sad, angry, surprised, disgusted, fearful
beforehand for introducing
the different features designed in the visual programming tool,
and also served as a basic
emotion command corpus for participants to use during live animation.
We adopted a laptop with Intel CPU i7 2.6GHz, 8 GB RAM, Nvidia GeForce 920M GPU
as the server to run the live animation system,
and also used the same computer as the audience side.
A laptop with Intel CPU i7 2.2GHz, 16 GB RAM, Intel Iris Pro GPU
was used as the performer side for participants to edit the emotion commands
and conducted live animation.

\subsection{Procedure}

An overview of our live animation system was introduced first (about 20 minutes),
including how to edit an emotion command, how to use the live animation function, and
the main potential usage scenario of the system: telepresence agents / robots.
Participants were then told to try and explore the system, and we would elaborate the unclear
functionalities they asked during the period (about 10 minutes).
Then we asked participants to edit four emotion commands, including the semantic meaning of
shy, excited, the feeling of hungry, and one that created by participants themselves.
No time limitation was set for this stage. After they finished, we conducted an interview to
ask their experience of the usability, good, and bad parts of our system.
With the four personal edited emotion commands and the six ones prepared beforehand,
participants were asked to use the live animation function to tell a self-chosen story
to the audience side
\footnote{The audience side has no human partner for interactive communication.
For the current evaluation, we only consider
using the live animation system from the performer side, and allow users to free explore its
functionalities.}.
Again, no time limitation was set. Finally, we collected the user experience on usability,
good, and bad parts of our live animation function through an exit interview.

\subsection{Results and Discussion}

In general, participants found the emotion command editing function easy-to-use, flexible for creating
diverse emotion expressions, and the live animation function novel, playful.

\subsubsection{Usability}
Participants found that the workflow and interface design for
editing an emotion command was intuitive, and fluent to create an emotion expression.
On average, participants took about 3:39 (minutes:seconds, min 1:13, max 9:17)
for editing one emotion command, as shown in Table
\ref{The cost time of editing an emotion command for each participant.}.
\emph{"I think the emotion command function is very interesting to use, and suitable for novices
to pick up the concept. It is also very flexible to construct an emoji with different components.
The action nodes provide several options to select different levels of motion, which is also quite good."
--P1.}

In addition, participants generally reflected that the live animation was interesting to use, and
acknowledged that the emotion command feature can enhance the expressiveness of emotions.
Participants also confirmed that the recommendation algorithm was useful, but not very accurate.
We elaborate the recommendation algorithm more in the application section.
\emph{"I think the live animation function is good to use. And the emotion command can help me express
interesting emotion effects, which is helpful for enhancing the emotional communication.
The recommendation panel is also helpful, but it is not very accurate currently."--P2.}
\emph{"If there are too many emotion commands, I will have no idea how to find a proper one
during live animation. Recommendation is apparently helpful. It is better if you can
make it more accurate."--P3.}

\begin{table}[h]
  \centering
  \begin{tabular}{ccccc}
    & Shy & Excited & Hungry & Free-form (semantic) \\
    \hline
    P1 & 2:40 & \textbf{1:13} & 1:55 & 2:10 (laugh) \\
    P2 & 6:32 & 1:57 & 3:08 & 3:22 (all blue) \\
    P3 & 4:01 & 4:15 & 3:00 & 3:11 (amazing) \\
    P4 & 2:15 & 6:01 & 1:48 & 4:18 (furious) \\
    P5 & \textbf{9:17} & 5:06 & 5:33 & 5:19 (I want you) \\
    P6 & 3:29 & 2:15 & 2:42 & 2:11 (exhausted)
  \end{tabular}
  \caption{The cost time of editing an emotion command for each participant(the shortest and longest
  times are highlighted).}
  \label{The cost time of editing an emotion command for each participant.}
\end{table}

\vspace{-20pt}

\subsubsection{Creativity}
\begin{figure}[h]
  \centering
    \includegraphics[width=1.0\columnwidth]{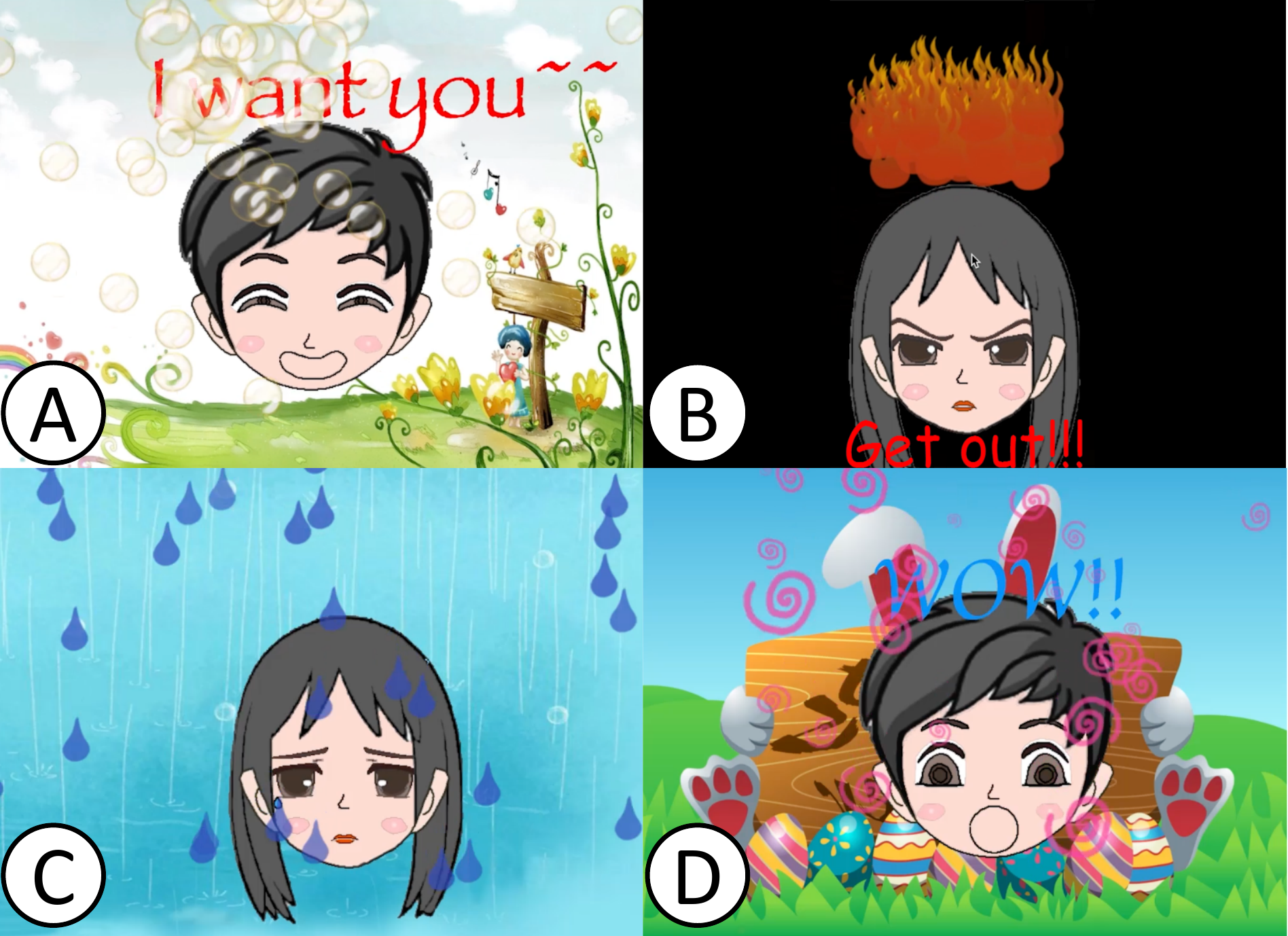}
  \caption{User defined emotional expressions: A) I want you (P5), B) furious (P4),
  C) all blue (P2), D) amazing (P3).}
  \label{The personal defined emotion expression}
\end{figure}

We asked participants to edit a personal emotion command with meaning defined by
themselves to let them freely explore the emotion command tool. Participants found the organized
way with the visual programming tool was quite useful for boosting their creativity,
especially for people without previous experience of graphics design.
Participants were generally satisfied with the basic features supported by our system.
In the free-form creation stage, participants authored a range of interesting
emotion expressions. Several examples are shown in
Figure \ref{The personal defined emotion expression}.
\emph{"It is good to have different components for me to compose an emotion expression. I can construct
it like building blocks. The provided danmaku node is really useful for conveying some subtle emotion meanings,
especially when the facial expression is not diverse enough."--P5.}
\emph{"With the visual programming tool, the logic is quite clear for me, and thus it is helpful for creating new
things. I like the function to change the background image. In this way, I can build more vivid emotion illustration."--P2.}

\subsubsection{Applicability}

As mentioned previously,
our potential target application scenario is for
telepresence agents or telepresence robots with a physical body.
We asked participants to use the live animation function with the mindset of
the target application scenario, and collected their experience of using it.
In general, the feedback is positive. Participants reported that
the function was playful and interesting, expressing emotions during live animation is also meaningful.
\emph{"Binding a key in the keyboard with the semantic meaning defined by myself is convenient to use.
With the edited emotion commands, I could use more expressive ways during communication."--P1.}

Participants thought the recommendation function was helpful, especially when there were
lots of emotion commands.
However, in the first trial of building our recommendation algorithm (using only Euclidean distance
to measure the similarity),
participants were not quite satisfied with the recommendation results.
We refined the algorithm with the separation of positive and negative emotions presented in the
system section, and manually fine-tuned the parameters of estimating the valence and arousal levels
to better align with the mental values self reported in the emotion commands.
In addition, we optimized the performance of live animation by pausing the computational intensive
emotion recognition algorithm
during emotion command triggering.
We asked two participants (P2, P3) that were not satisfied with the recommendation
function in the previous experiment
to re-try our live animation feature, and
they found the improved version was much better than the previous one.
\emph{"I find the recommendation becomes accurate, and the live
animation is also more fluent than the previous one."--P2.}
\emph{"It is better now. Although I cannot say it is perfect. But separating positive and negative
emotions can improve the using experience."--P3.}


\subsubsection{Interesting Findings} During our user evaluation, participants also provided
some interesting suggestions to improve our system, which inspires us to conduct 
further works to improve the system. 

\textbf{Generating emotion commands}. 
One participant suggested it could be useful if the system
can recognize users' emotions, and generate an emotion command automatically to serve as an initial 
template.
\emph{"If possible, it would be good if the system can recognize my emotion, and generate an emotion command
automatically. And then the generated emotion command can serve as a template so that it may better
fit my expectation."--P5.}

\textbf{Combining emotion commands}. Due to the limited number of keyboard devices,
the number of executable emotion commands is limited. Participants suggested it is 
possible to combine several emotion commands to generate a new emotional expression.
In addition, in this way, it expands the emotion vocabulary for users. 
\emph{"It would be interesting if I can trigger two emotion commands simultaneously, and the
system shows the combined emotion expression. I think, in that way,
it can give me more flexibility to represent subtle emotions"--P2}.

\section{Conclusion}
In this paper,
we explore the emotional expressiveness issue of 2D live animation. In particular,
we propose a descriptive emotion command model for binding a triggering action,
the semantic meaning, psychology measurements, and behaviors. Based on the model,
we design and implement a proof-of-concept 2D live animation system
for scenarios like telepresence agents or robots for peer-to-peer communication.
Through a preliminary user evaluation, we showcase the usability of our system, and
indicate its potential for boosting creativity and enhancing the emotional communication experience.
In the further works, we plan to
enrich the functionalities through computer vision techniques, and also explore its usage
in real world applications.

\bibliographystyle{ACM-Reference-Format}

\end{document}